\documentclass[a4paper,11pt]{article}
\pdfoutput=1 

\usepackage{jinstpub} 
\usepackage{multirow}

\usepackage[version=3]{mhchem}
\usepackage{siunitx}
\usepackage{booktabs}
\usepackage[numbers]{natbib}

\title{Molecular sieve vacuum swing adsorption purification and radon reduction system for gaseous dark matter and rare-event detectors}

\author[a,b,1]{R.R. Marcelo Gregorio,\note{Corresponding author.}}
\author[b]{N.J.C. Spooner}
\author[c,d]{F. Dastgiri,}
\author[b]{A. C. Ezeribe,}
\author[c,d]{G. Lane,}
\author[b]{A.G. McLean,}
\author[e]{K. Miuchi,}
\author[f]{and H. Ogawa}

\affiliation[a]{School of Physical and Chemical Sciences, Queen Mary University of London, E1 4NS, United Kingdom}

\affiliation[b]{Department of Physics and Astronomy, University of Sheffield, South Yorkshire, S3 7RH, United Kingdom}

\affiliation[c]{Department of Nuclear Physics and Accelerator Applications, Australian National University, Australia}

\affiliation[d]{ARC Centre of Excellence for Dark Matter Particle Physics, Australia}

\affiliation[e]{Department of Physics, Kobe University, 1-1 Rokkodai-cho, Nada-ku, Kobe, Hyogo, 657-8501, Japan}

\affiliation[f]{CST Nihon University, Surugadai, Kanda, Chiyoda-ku, Tokyo, 180-0011, Japan}

\emailAdd{robert.gregorio@qmul.ac.uk}

\abstract{In the field of directional dark matter experiments \ce{SF6} has emerged as an ideal target gas. A critical challenge with this gas, and with other proposed gases, is the effective removal of contaminant gases. This includes radon which produce unwanted background events, but also common pollutants such as water, oxygen and nitrogen, which can capture ionisation electrons, resulting in loss of detector gas gain over time. We present here a novel molecular sieve (MS) based gas recycling system for the simultaneous removal of both radon and common pollutants from \ce{SF6}. The apparatus has the additional benefit of minimising gas required in experiments and utilises a Vacuum Swing Adsorption (VSA) technique for continuous, long-term operation. The gas system’s capabilities were tested with a 100 L low-pressure \ce{SF6} Time Projection Chamber (TPC) detector. For the first time, we present a newly developed low-radioactive MS type \(5\text{\AA}\). This material was found to emanate radon at \(98\%\) less per radon captured compared to commercial counterparts, the lowest known MS emanation at the time of writing. Consequently, the radon activity in the TPC detector was reduced, with an upper limit of less than  \(7.2 \, \text{mBq}\) at a \(95\%\) confidence level (C.L.). Incorporation of MS types \(3\text{\AA}\) and \(4\text{\AA}\) to absorb common pollutants was found successfully to mitigate against gain deterioration while recycling the target gas.}


\arxivnumber{1234.56789} 

\begin{document}

\maketitle

\flushbottom

\section{Introduction}
\label{sec:intro}
As next-generation direct detection dark matter experiments increase in sensitivity, they approach an irreducible neutrino background \cite{billard2014implication}. If no dark matter signals are detected, this background establishes an ultimate discovery limit for these experiments. Directional detectors offer a way to mitigate this issue, as they can discriminate against the neutrino background by providing additional information on the direction of the nuclear recoils induced by events \cite{vahsen2021directional}. Various approaches to directional detectors exist \cite{golovatiuk2021directional, alexandrov2021directionality}, but gas-based Time Projection Chambers (TPCs) are the most commonly employed \cite{mayet2012drift, santos2011mimac, CYGNO2023gud}. These detectors use low-pressure gas targets, which produce significantly longer nuclear recoil tracks than those generated in higher-density target media. \ce{SF6} has gained popularity in gas-based directional detectors due to its novel properties \cite{ellis1991elastic, Phan_2017}. Recent developments support \ce{SF6} as the target gas for such searches \cite{McLean2023dnh, Eldridge2023zuy, Higashino2023mma}.

Two independent problems arise in the operation of gas-based directional dark matter detectors due to contaminant gases. The first is radon contamination, which originates from the intrinsic radioactive background of the detector material. The second involves common pollutants, such as water, oxygen, and nitrogen, introduced by outgassing and leaks. Radon contamination can act as a source of unwanted background noise, mimicking genuine signals. This occurs when radon progeny undergo alpha decay towards the detector walls, leading to a slowly recoiling lead nucleus entering the detector volume. This nucleus can then interact with the target medium, producing a nuclear recoil signal similar to that of a WIMP \cite{Battat_2014}. Note that in rare-event physics experiments, \(\ce{^222Rn}\) is the primary isotope of concern for radon contamination due to its abundance and the longest half-life of 3.8 days among radon isotopes. Therefore,  any reference to radon in this work corresponds to this to \(\ce{^222Rn}\).

Common pollutants can reduce the detector's amplification capabilities by capturing electrons produced during interactions \cite{guida2020effects}. These are two separate issues, however this work explores the possibility that both can be addressed using a single new gas purification system with appropriate Molecular Sieve (MS) filters.

To effectively incorporate these filters into gas-based directional dark matter detectors, a gas system utilising Vacuum Swing Adsorption (VSA) technology, complemented by a gas recovery buffer, was designed. The VSA technique enables on-site regeneration of the MS filters, a method already employed in various physics experiments, typically in activated-charcoal-based gas systems for low radon clean rooms \cite{ pocar2005low, https://doi.org/10.48550/arxiv.1404.5811, street2018radon}. A key innovation here is the addition of the gas recovery buffer. This component captures the small volume of gas lost during the regeneration phase of conventional VSA systems. By doing so, it maximises the amount of recycled gas. This is particularly crucial for detectors that rely on fluorine-containing gases, such as \ce{SF6} that pose a risk to the environment.

To assess the performance of the gas system, a prototype was constructed and applied to a lab-based Time Projection Chamber (TPC) detector with Thick Gas Electron Multiplier (ThGEM). Details of this and of the newly developed low-radioactivity MS are discussed in \autoref{Chapter:GasSystemDemo:setup}. Performance tests were conducted both with and without the gas recirculation system in place. Assessment of the reduction of intrinsic radon contamination in the detector setup is detailed in \autoref{Chapter:GasSystemDemo:rgr}, while evaluations concerning the preservation of the detector's gain amplification capabilities are detailed in \autoref{Chapter:GasSystemDemo:ggr}.

\section{ThGEM-based TPC detector with gas system setup}
\label{Chapter:GasSystemDemo:setup}
\begin{figure}[ht]
\centering 
\includegraphics[height=7.5cm]{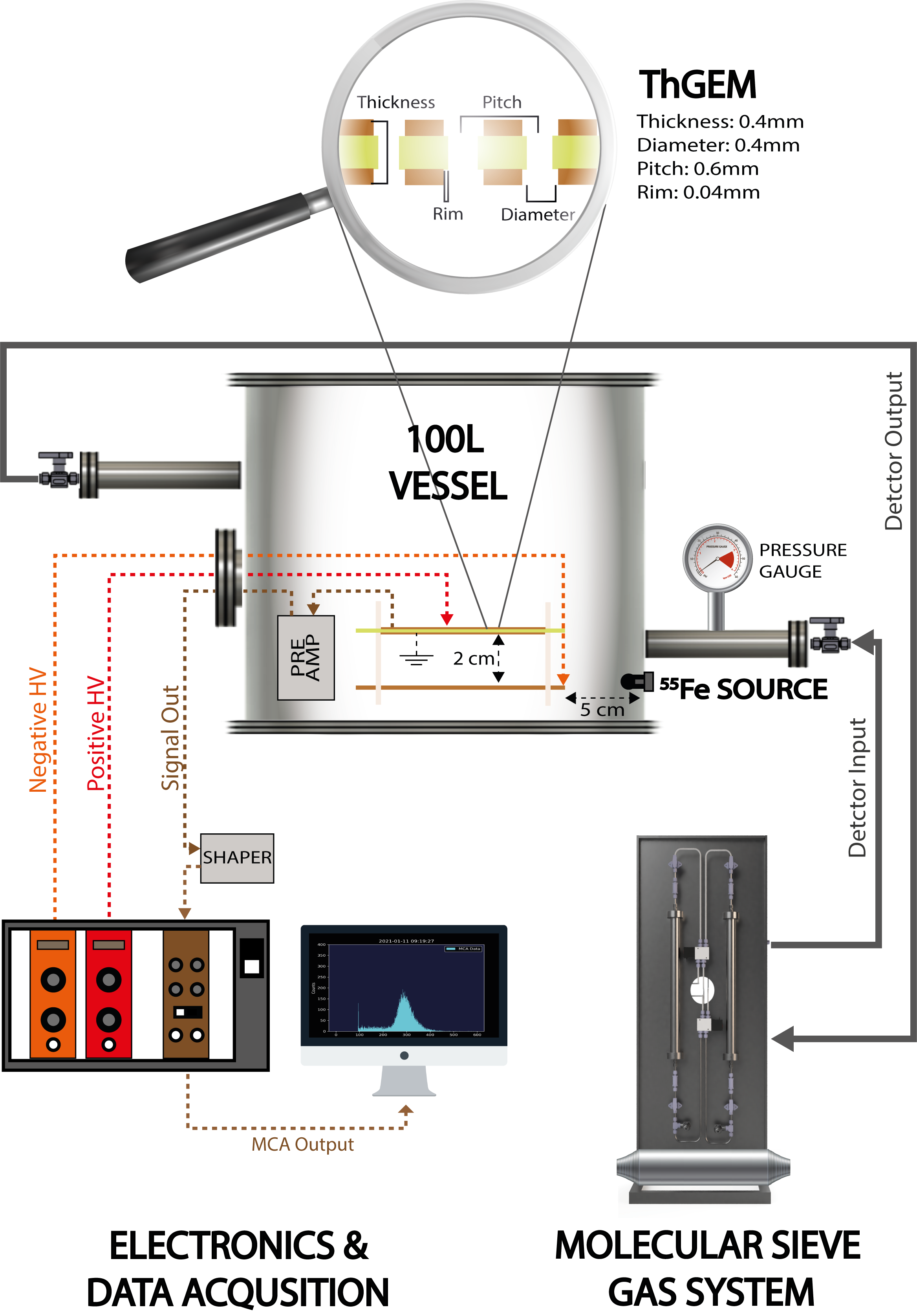}
\caption{\label{fig:demosetup} Schematic of the experimental setup used in the gas system performance testing with a ThGEM-based TPC detector.}
\end{figure}

\autoref{fig:demosetup} shows the experimental setup used to assess the performance of the gas system. The main components of the setup are the TPC detector which has a ThGEM readout, electronics \& Data Acquisition, and the gas system prototype including the MS filter.

 A ThGEM readout was employed because it has been utilised in a large number of gain studies in the past \cite{baracchini2018negative, ezeribe2021demonstration, breskin2009concise}. Gain is a measure of the detector's amplification capabilities and is the parameter tested in \autoref{Chapter:GasSystemDemo:ggr}, to demonstrate that the gas system is common pollutants detrimental to gas gain. For the performance tests, continuous long-term detector operation is required, so the ThGEM and TPC configuration chosen was based on previous work which has demonstrated stable operation \cite{eldridge2021new, scarff2017developments}. The ThGEM used was 10$\times$10 cm with thickness  0.4 mm, hole pitch 0.4 mm, and a hole diameter 0.4 mm (see top of \autoref{fig:demosetup}. The ThGEM detector was mounted 2 cm from a square cathode to create a time projection chamber. To achieve the electric field required to drift electrons, high-voltage power supplies were connected to the cathode and the top of the ThGEM, and the bottom of the ThGEM was grounded, as shown by the dashed lines in \autoref{fig:demosetup}. The high-voltage power supplies used were Bertan model 377P for the positive supply and Bertan model 377N for the negative supply. The signal was read out from the top of the ThGEM via an Ortec 142 IH preamplifier, connected to an Ortec 572 shaping amplifier, with signals recorded via an Ortec 926 ADCAM MCB in the form of a pulse height spectrum on a computer, where the gain was subsequently calculated. To provide a standard source of ionisation in the TPC, an \ce{^55Fe} source producing 5.89 keV X-rays was mounted on a magnet and directed at the sensitive detector volume, as shown to the right of the 100 L vessel. The magnet allowed the source to be redirected for source-off measurements. Although the sensitive detector volume is only 0.2 L, the detector was enclosed in a 100 L vessel to demonstrate the gas system's capability with large volumes, which is important for future large-scale TPCs. It should be noted that, with the exception of the DRIFT experiment, most directional dark matter gas TPCs are less than 50 L \cite{mayet2012status, nakamura2015newage, CYGNO:2023ucc}.

\begin{figure}[h]
\centering 
\includegraphics[height=7.5cm]{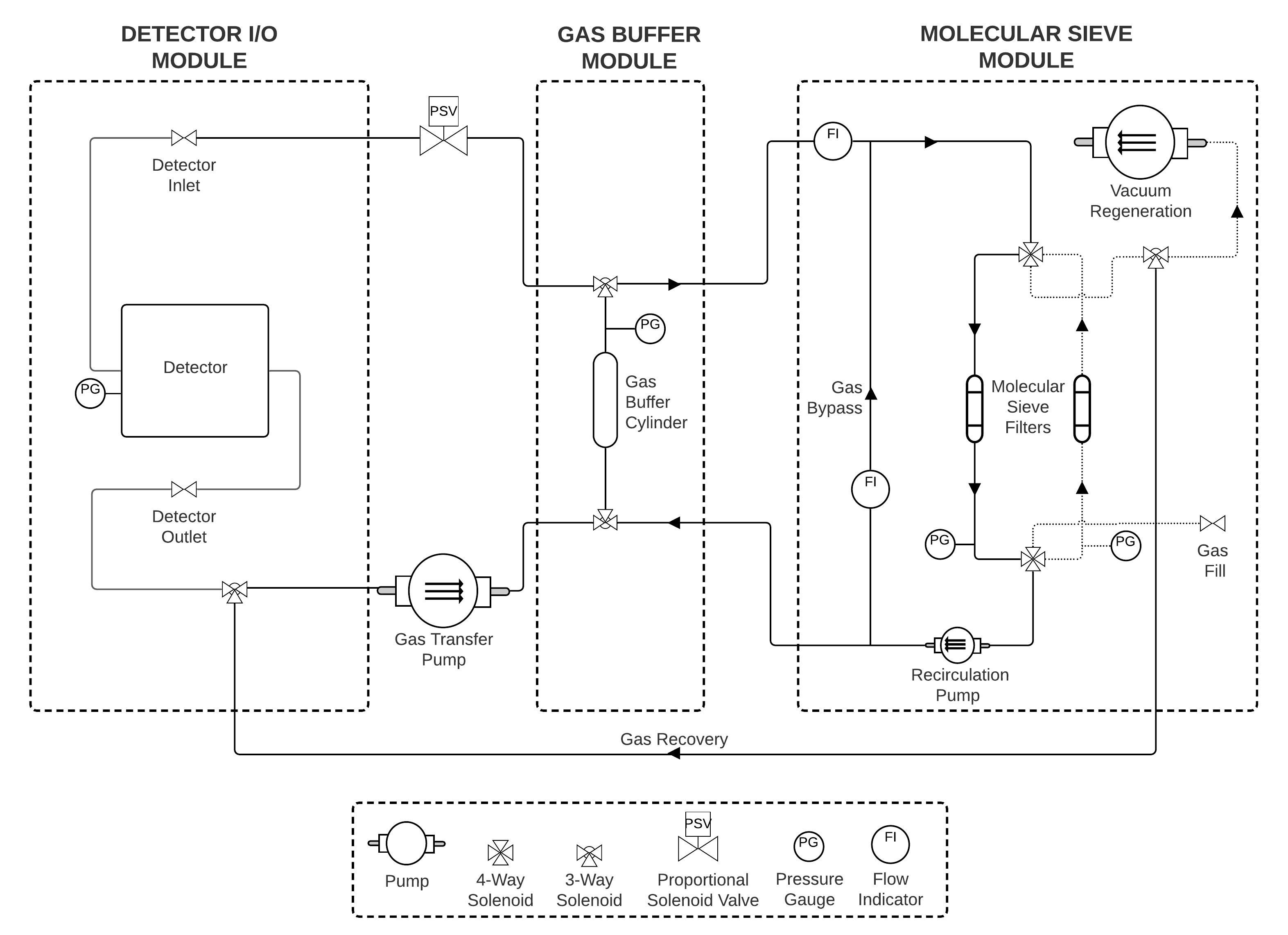}
\caption{\label{fig:gs_detailed} Schematic of the experimental setup used in the gas system performance testing with a ThGEM-based TPC detector. Note the gas recovery line is used to feed lost gas back into the buffer.}
\end{figure}

The molecular sieve gas system prototype is shown on the right of \autoref{fig:demosetup}, connected to opposing arms of the vacuum vessel to optimise gas flow. Details of the gas system design are shown in the Piping and Instrumentation Diagram in \autoref{fig:gs_detailed}. It is structured into three primary modules: (1) The Molecular Sieve Module, which includes two MS filters with a capacity of up to 500 g and four-way solenoid valves for efficient gas routing. (2) The 4.5 L Gas Buffer Module, which serves as an intermediary, temporarily storing the gas during filtration and subsequently refilling the detector with clean gas. (3) the Detector Input/Output (I/O) Module, which controls the inflow and outflow of gas in the detector vessel, utilising a proportional solenoid valve and a gas transfer pump to regulate pressure levels. 

\begin{figure}[h]
\centering 
\includegraphics[height=7.5cm]{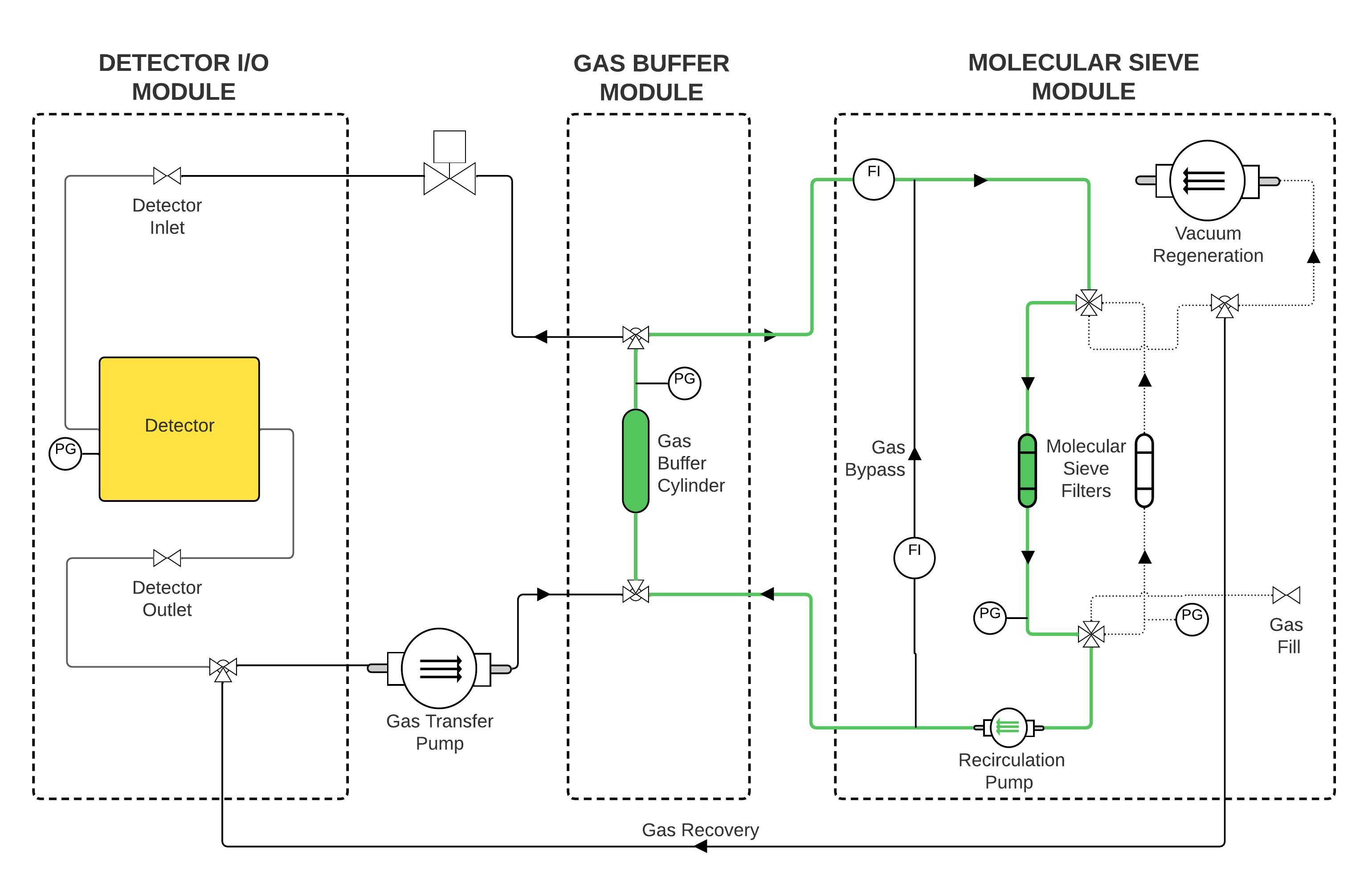}
\caption{\label{fig:GASSYSTEM_OPERATIONVOLS} Schematic of the two separate volumes within the system: the gas inside the detector (shown in yellow) and inside the buffer (shown in green). The green loop corresponds to filtration using the gas buffer cylinder, while the gas shaded in yellow is the gas during detector operation. }
\end{figure}

The operation of the gas system can be understood by considering two separate volumes, the gas inside the detector vessel and gas inside the buffer, indicated in \autoref{fig:GASSYSTEM_OPERATIONVOLS} as yellow and green areas, respectively. While the detector is online, the buffer volume is continuously filtered by the MSs. After a certain period, the detector volume will become contaminated by common pollutants and radon emanation. To combat this, the cleaner gas inside the buffer is flowed to the detector vessel, effectively cleaning the detector gas by dilution. The detector is restored to its original pressure by transferring the additional gas to the buffer and resumes filtration. Detector volumes are generally larger than the gas buffer volume, so the pressure in the buffer is higher. For example, for this work, the pressure in the detector was \( 50 \, \text{torr} \), and in the buffer \( 1.2 \times 10^{3} \, \text{torr} \).

\begin{figure}[h]
\centering 
\includegraphics[height=7.5cm]{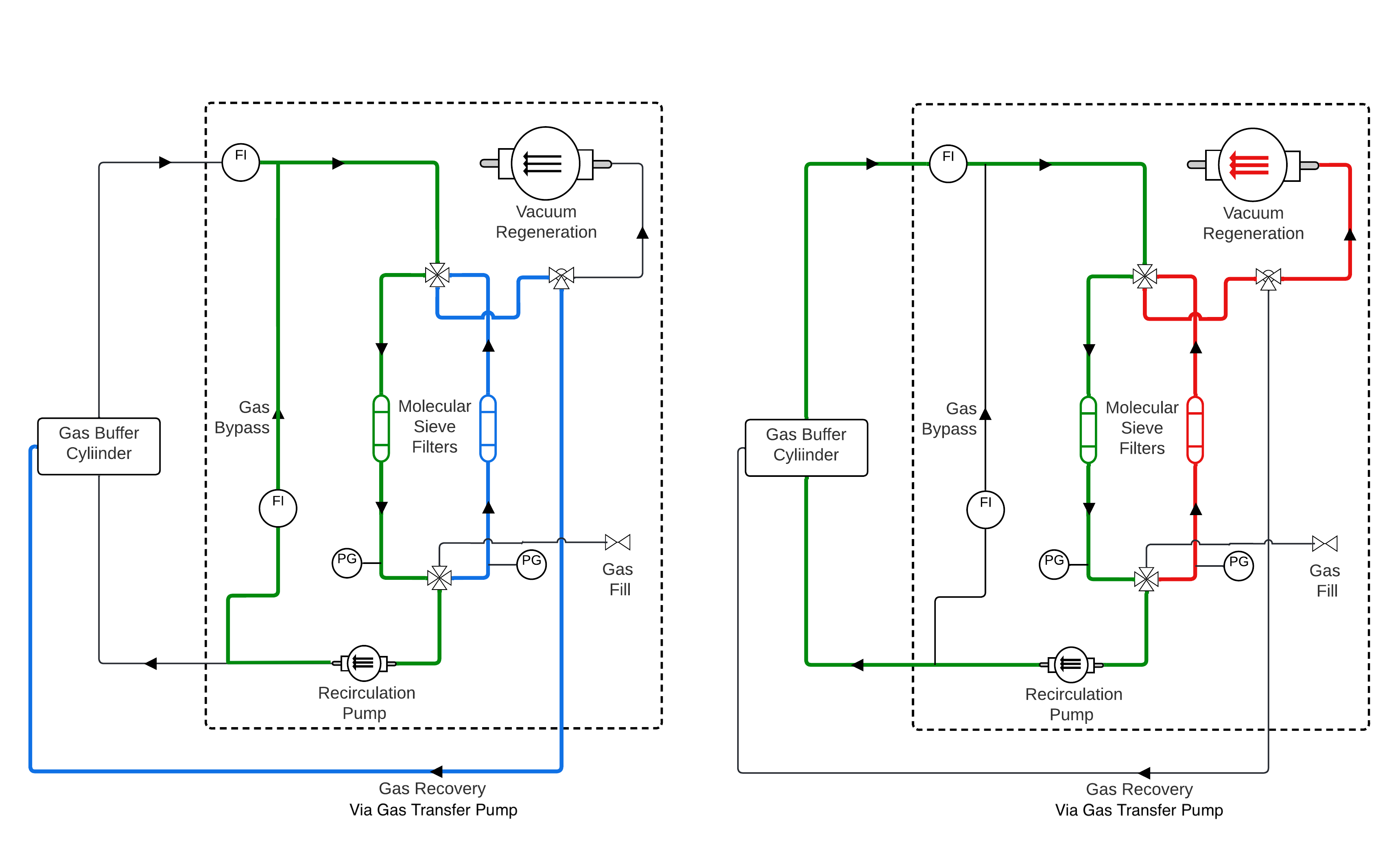}
\caption{\label{fig:GASSYSTEM_FILT_VACGEN} Schematic of the molecular sieve module during gas recovery (left) and vacuum regeneration (right).}
\end{figure}

The dual MS column configuration ensures an MS filter is always available by allowing simultaneous filtration and regeneration. The process of filtration and regeneration is schematically shown in \autoref{fig:GASSYSTEM_FILT_VACGEN}. The filtration process (green line) is achieved by continuous gas flow through the MSs and gas buffer, driven by a recirculation pump. The regeneration process occurs in two steps, gas recovery and vacuum regeneration. Gas recovery (blue line) corresponds to the collection of the small gas volume in the MS filters, which is lost in conventional VSA during vacuum regeneration. The gas is collected by evacuating the MS filter using the gas transfer pump, with the output redirected to the gas buffer cylinder. The MS filter is evacuated just above the critical regeneration pressure, $\mathcal{O}$(10 torr) \cite{doi:10.1021/acs.iecr.8b00798, Arthurs_2021}, ensuring that most of the gas is recovered whilst avoiding the release of captured contaminants. Once the filter gas is recovered, vacuum regeneration (red line) is initiated by applying a sub-torr vacuum.

\begin{figure}[h]
\centering 
\includegraphics[height=5.5cm]{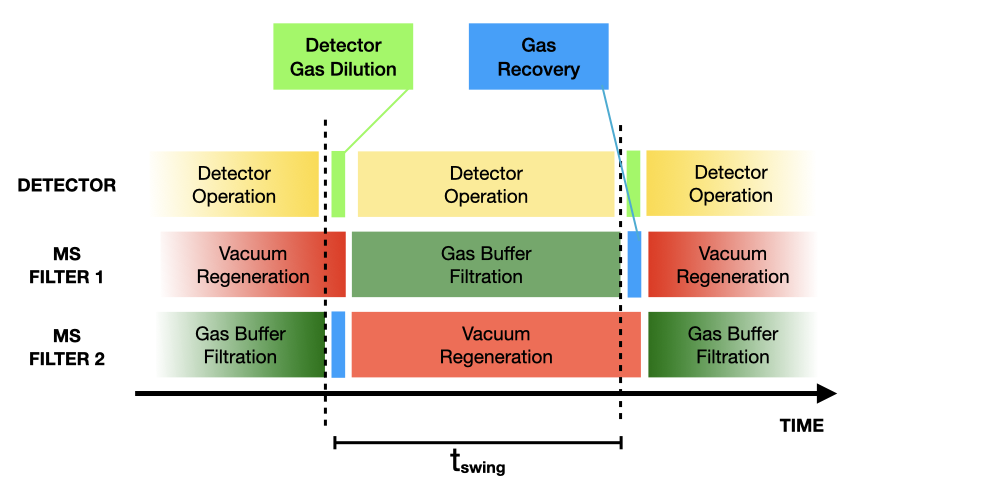}
\caption{\label{fig:GASSYSTEM_TIMELINE} Example timeline of VSA operation showing the operation modes for the detector and dual MS filters. Here, \(t_{\text{swing}}\) is in the order of days, and detector gas dilution and gas recovery are in minutes.}
\end{figure}

\autoref{fig:GASSYSTEM_TIMELINE} shows an example of the operational timeline of the dual MS filters and the detector. Here, \(t_{\text{swing}}\) is the time between MS swings, which coincides with gas dilution. it is essential to distinguish between these switching processes to understand how the gas system operates. The MS swing involves switching between MS filter 1 and MS filter 2 to transition between filtration and regeneration, and vice versa. Gas dilution effectively switches the gas between the filtered gas in the buffer and the used gas in the detector. The MS swing must be done within the breakthrough time of the filter to avoid the accumulation of contamination. The time for breakthrough is a function of the MS filter's dimensions and flow parameters, which must be calibrated for operation. The required frequency for the gas dilution process depends on the detector's contamination rate. In this work, the timescale for detector gas dilution and gas recovery is minutes, whereas detector operation, filtration, and regeneration span over days.

Following  publication of the initial study identifying the properties of low-radioactive MS candidates \cite{Marcelo_Gregorio_2021}, an updated version known as NU-V2 MS was developed by Nihon University. Although the original version exhibited up to 61 ± 9\% lower radon emanation per radon captured compared to commercially available MS, it was suspected that the calcium ion used in its synthesis was the primary source of radon emanation. To enhance the performance further, a cleaner calcium supplier was used for the NU-V2 MS \cite{ogawa2022measurement}.

To ensure a fair comparison, the methodology outlined in the original paper was followed. This involved evaluating both the intrinsic emanation and radon filtration capabilities of the molecular sieves to derive a parameter for radon emanated per radon captured. The results for both the original NU MS and the new NU-V2 MS are presented in \autoref{tab:NUV2}.

\begin{table}[ht]

\centering
\smallskip
\begin{tabular}{|c|c|c|c|}
\hline
\begin{tabular}[c]{@{}c@{}} \textbf{NU-developed} \\ \textbf{MS}\end{tabular} 
&\begin{tabular}[c]{@{}c@{}} $^{222}$\textbf{Rn Captured} \\ \textbf{ per kg (Bq kg$^{-1}$})\end{tabular}  & 
\begin{tabular}[c]{@{}c@{}} $^{222}$\textbf{Rn Emanated} \\ \textbf{ per kg (mBq kg$^{-1}$})\end{tabular}  &
\begin{tabular}[c]{@{}c@{}} $^{222}$\textbf{Rn Emanated per} \\ \ $^{222}$\textbf{Rn Captured ($\times 10^{-3}$)}\end{tabular} \\
\hline
V1 (Granules) & 35±2 & 99±23 & 2.8±0.7 \\
V1 (Powder) & 330±3 &  680±30 & 2.1±0.1 \\
V2 (Powder) &  254±3 & <14.4  & <5.7$\times 10 ^{-2}$ \\
\hline
\end{tabular}
\caption{\label{tab:NUV2} Radon filtration, intrinsic MS emanation and comparison parameter results for the NU-developed MS in granule and powdered form and NU-developed MS V2.}
\end{table}

The ideal MS should exhibit both a high radon capture rate and low intrinsic emanation. While the capture rate can be enhanced by increasing the surface-to-volume ratio, for example by converting granules to powder, as demonstrated in the comparison between V1 powder and V1 granules, this approach has the drawback of facilitating easier emanation of radon from the MS material. A balanced strategy is needed to optimise capture rate while minimising radon emanation.

The geometry of the NU-developed MS (V2) closely resembles that of its powdered counterpart, NU-developed MS (V1), as evidenced by their similar radon reduction efficiency measurements: \(254 \pm 3\) Bq kg\(^{-1}\) for V2 and \(330 \pm 3\) Bq kg\(^{-1}\) for V1. The intrinsic radon emanation of V2 was measured with an upper limit of \(14.4 \, \text{mBq kg}^{-1}\) at a 95\% C.L., significantly lower than the \(680 \pm 30 \, \text{mBq kg}^{-1}\) measured for V1. The larger error in the V2 emanation measurement is expected as it approaches the experimental setup's background limits. This comparison clearly demonstrates the effectiveness of the new calcium ion supplier in reducing the MS overall emanation.

The upper limit for the radon emanated per radon captured parameter is \(5.7 \times 10^{-5}\), representing at least a \(98.9\%\) reduction in radon emanated per radon captured compared to the commercial Sigma-Aldrich MS. To the author's knowledge, these are the lowest intrinsic radon emanation rates per unit mass for any molecular sieves. NU-V2 MS will be integrated into the gas system for the radon reduction test discussed in section \autoref{Chapter:GasSystemDemo:rgr}.

\section{Radon activity reduction test}
\label{Chapter:GasSystemDemo:rgr}
To investigate the gas system's impact on radon mitigation in a ThGEM-based TPC detector, understanding radon dynamics between the gas system and the TPC is crucial. Equilibrium radon activity depends on the balance between emanation from materials and absorption in the MS filter

The radon activity due to materials in the ThGEM-based TPC detector volume and gas system can be described by \autoref{eqn:radonema_TPC} and \autoref{eqn:radonema_GS}, respectively. 
\begin{equation}
     A^{\text{ema}}_{\text{TPC}}(t_{\text{ema}}) = A^{\text{sec}}_{\text{TPC}} - \left(A^{\text{sec}}_{\text{TPC}} - A_{\text{TPC}}(t_{\text{ema}} = 0)\right) \exp(-\lambda_{\text{Rn}} t_{\text{ema}}),
\label{eqn:radonema_TPC}
\end{equation}
\begin{equation}
     A^{\text{ema}}_{\text{GS}}(t_{\text{ema}}) = A^{\text{sec}}_{\text{GS}} - \left(A^{\text{sec}}_{\text{GS}} - A_{\text{GS}}(t_{\text{ema}} = 0)\right) \exp(-\lambda_{\text{Rn}} t_{\text{ema}}),
\label{eqn:radonema_GS}
\end{equation}

Here, \( A^{ema} \) is the radon activity due to emanation, \( t_{ema} \) is the emanation time, \( A^{ema}(t_{ema} = 0) \) is the initial radon activity at zero emanation time, $A^{\text{sec}}$ is the activity at secular equilibrium and \( \lambda_{\text{Rn}} \) is the radon decay constant. The subscripts TPC and GS correspond to the origin of radon emanation; for instance, \( A^{ema}_{\text{GS}} \) is the radon activity from material emanation in the gas system volume.

Radon adsorption on MS during filtration is modelled kinematically, incorporating the molecular flux \( F \), sticking probability \( S \), and filtration time \( t_{\text{filt}} \) \cite{LibreTextChem2022}. The incident molecular flux \( F \) can be described using the Hertz-Knudsen relation, which is a function of gas pressure \( P \), molecular mass of absorbent species \( m \), temperature \( T \), and the Boltzmann constant \( k \) \cite{kolasinski2012surface}.  Given constant gas system operation conditions, \( F \) can be approximated to be directly proportional to the total radon atom count \( N_{tot} \). The sticking probability \( S \) is described by

\begin{equation}
    S = f(\theta) \exp\left(-\frac{E_a}{RT}\right),
    \label{eqn:radonfiltS}
\end{equation}

here $f(\theta)$ is a function related to the surface coverage of adsorbed species on the MSs, $E_a$ is the activation energy barrier for adsorption, and $R$ is the gas constant. Since the MSs are regularly vacuum regenerated it can be approximated that there are always vacant sites, additionally if the same MS geometry is used the number of available sites remain the same. Therefore, a reasonable first approximation is that the radon sticking probability, $S$ is constant when operating with the gas system as $f(\theta)$ is expected to remain relatively unchanged. Using the assumptions discussed above, the number of radon atoms captured by MS during filtration can be estimated by
\begin{equation}
    N_{\text{MS}} \approx N_{\text{tot}} \times k_{\text{ms}} \times t_{\text{filt}},
    \label{eqn:radonfiltermodelapprox}
\end{equation}

where $k_{ms}$ is a constant associated with the sticking probability for a fixed MS geometry and parameters related with the incident molecular flux, such as pressure, flow rate, and temperature, which are assumed to remain constant during normal gas system operation. The number of radon atoms captured can be converted to activity by using  $N=A/\lambda_{Rn}$, where $\lambda_{Rn}$ is the radon decay constant, resulting in  
\begin{equation}
    A_{MS} \approx A_{tot} k_{ms} t_{filt},
    \label{eqn:radonfilt}
\end{equation}
here $A_{MS}$ is the captured activity due to the radon adsorbed by the MSs and $A_{tot}$ is the radon total activity.

Recall that the parameter \(t_{\text{swg}}\) is the time set between the gas dilution and swing process. Therefore, for every swing cycle, the gas system volume and TPC volume remain separate for a duration of \(t_{\text{swg}}\). During this time, the radon activity in the TPC volume, $A_{TPC}$, and gas system volume $A_{GS}$ can be described by \autoref{eqn:TPCdswing} and \autoref{eqn:GSdswing}, respectively.
\begin{equation}
  A_{\text{TPC}} = A^{\text{ema}}_{\text{TPC}}(t_{\text{swg}}),
  \label{eqn:TPCdswing}
\end{equation}
\begin{equation}
  A_{\text{GS}} = A^{\text{ema}}_{\text{TPC}}(t_{\text{swg}}) - A_{\text{MS}}(N_{\text{tot}}, t_{\text{swg}}).
  \label{eqn:GSdswing}
\end{equation}

The radon activity in the TPC volume is expected to increase due to material emanation from the detector setup, while the radon activity in the gas system is expected to decrease, assuming that the rate of radon filtration is greater than the rate of material emanation from the gas system.

After time $t_{swg}$, the dilution and swing process is initiated, and the gas system and TPC volumes are mixed. Despite the different sizes of the two volumes, there are approximately equal amounts of gas in each volume due to the pressure differences. Consequently, the resulting activity after the swing process for both volumes is approximately half of the combined activity from the gas system and TPC, assuming it is well mixed. Since gas system operation will involve many swings, the radon activity in the ThGEM-based TPC volume after $n$ cycles is given by
\begin{equation}
A_{V_{\text{TPC}}}(n) = \frac{1}{2}\sum_n \left( A^{\text{ema}}_{\text{TPC}}(t_{\text{swg}}) + A^{\text{ema}}_{\text{GS}}(t_{\text{swg}}) - A_{\text{MS}}(N_{\text{tot}}, t_{\text{swg}}) \right).
\label{eqn:radondynamicsmodelfull}
\end{equation}

\begin{figure}[hbtp]
\centering 
\includegraphics[width=7cm]{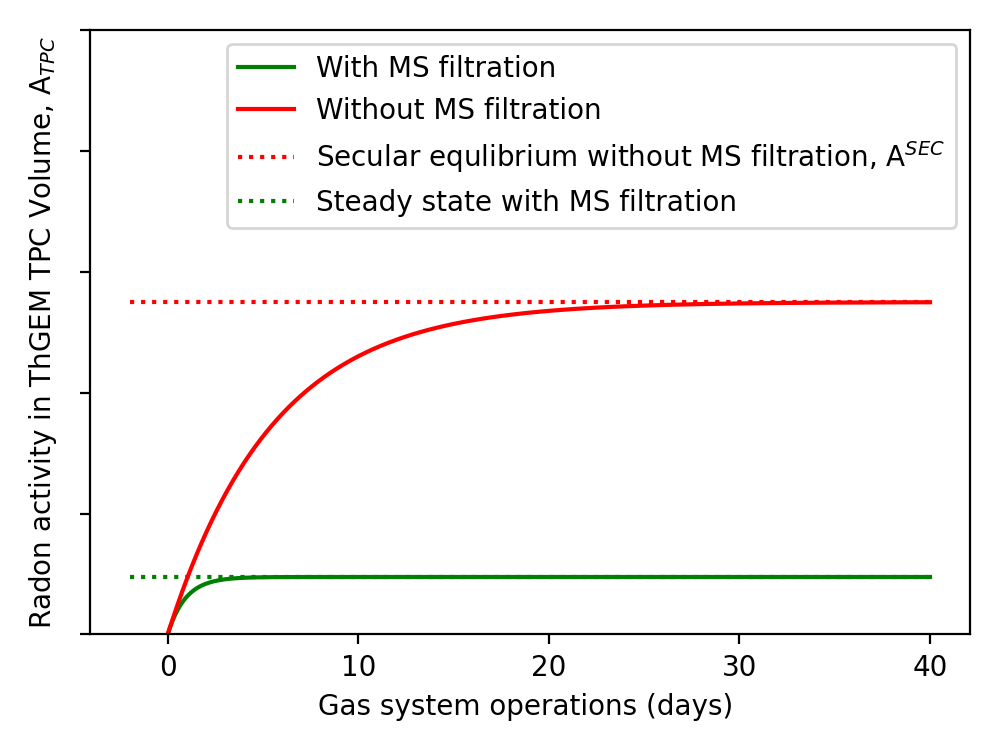}
\caption{\label{fig:radondynamicsplot} Model of radon dynamics in the TPC volume over 40 days operation, with and without MS filtration. Derived using \autoref{eqn:radondynamicsmodelfull}. }
\end{figure}

A comparison of radon dynamics within the TPC, with and without MS filtration, is presented in \autoref{fig:radondynamicsplot}. The model, using \autoref{eqn:radondynamicsmodelfull}, sets secular equilibrium activity parameters to unity and specifies the swing process interval, \( t_{swg} \), at 24 hours to align with the routine exchange of the detector volume. For the scenario involving MS filtration, \( k_{ms} \) is derived from radon filtration data as reported in \cite{Marcelo_Gregorio_2021}, with the value approximated to be \( 2 \times 10^{-5} \) s\(^{-1}\). Without MS filtration, \( k_{ms} \) is assumed to be zero. In both cases, initial radon activity in the TPC and gas system is considered as zero after thorough initial evacuation.

The behaviour of radon dynamics simulation can be described as follows. In the TPC, radon activity initially increases due to emanation from materials. As radon levels rise, the MS filter captures more radon, enhancing the filtration process. Emanation rates decrease over time as the system approaches secular equilibrium. A steady state is achieved when the radon emanation rate becomes equal to the radon filtration rate. The model predicts that this steady state is attained within a few days. In contrast, without MS filtration, radon activity increases until it reaches secular equilibrium over approximately one month, as is  expected. Here secular equilibrium activity can be considered to be the maximum radon emanation from materials during detector operation, and steady state activity corresponds to the reduced level achievable through gas system filtration.

To assess the gas system’s effectiveness in reducing radon activity in the gas TPC detector, it is necessary to compare the steady-state radon activity with MS filtration against the secular equilibrium activity without MS filtration. According to the radon dynamic model, operating the gas system with the TPC detector for four days is ample time to reach steady-state activity. Hence, a four-day run of the detector was performed with 40g of low-radioactivity MS type \(5\text{\AA}\) loaded in each filter of the gas system. The TPC detector was maintained at 50 torr \ce{SF6}, with a \(t_{\text{swg}}\) 24-hour interval for the swing and dilution cycle. This was followed by a 12-hour radon measurement to determine the radon activity in the TPC detector.

The radon activity measurement was conducted by the method of sampling with a DURRIDGE RAD7 radon detector \cite{instrumentation2017rad7}. Given that the RAD7 is calibrated to measure at atmospheric pressure, gas from the TPC's 100L volume was transferred into a smaller 4.5L sample cylinder and pressurised to atmospheric pressure. Two RAD7 detectors connected in a loop to the sample cylinder with internal pumps recirculating the gas were used for the 12-hour measurement.

To determine a value for secular equilibrium activity without filtration, an identical four-day operation was conducted with the gas system's filters removed, followed by a radon measurement. The additional time required to reach secular equilibrium is corrected for in the analysis. To account for any additional intrinsic background due to the radon activity measurement apparatus, the RAD7 detectors were purged with low-humidity low-radon \ce{SF6} before each measurement and a blank 12-hour measurement was conducted and used in the analysis.

The DURRIDGE RAD7 is not calibrated to measure in carrier gas \ce{SF6}, therefore a calibration factor must be applied to account for changes in the collection efficiency due to the carrier gas \ce{SF6}. To account for this, it was determined that the output of a DURRIDGE calibrated RAD7 must be multiplied by 3.33 \cite{InternalGasCorrection2023}. The RAD7 output is also multiplied by the total volume of the two RAD7s and sampling cylinder (6.4 L) in order to convert radon concentration output (Bq/m\(^3\)) to radon activity (Bq). Since radon measurements were taken using a sampling technique, corrections are required for the radon in the sample that has decayed during measurement and the radon contribution of the sampling apparatus. At the point of sampling, the source of radon is no longer present and will start to decay. Here the radon source is the intrinsic radon emanation of materials in the gas system and TPC setup. To account for the radon decay during the 12 hour measurement, the following equation is used:
\begin{equation}
    A(t_{\text{sam}}) = A_0 \exp(-\lambda_{\text{Rn}} t_{\text{sam}}),
    \label{eqn:sample_decay}
\end{equation}
where $A(t_{\text{sam}})$ is the radon activity at time since sampling, $A_0$ is the radon activity at the point of sampling, and $\lambda_{\text{Rn}}$ is the radon decay constant. The radon activity at the point of sampling $A_0$ is extrapolated from the equation fitted to the RAD7 radon activity data.

At the point of sampling, a new radon source is also introduced from the material emanation of the measurement apparatus. In order to account for this, background subtraction is applied to the extrapolated radon activity $A_0$. The background activity was calculated by conducting a blank 12 hour test using the radon measurement apparatus. The blank \(\ce{SF6}\) measurement resulted in an activity contribution of \(14.0 \pm 5.7\) mBq. A further correction is necessary for measurement without MS filtration because, at the time of sampling, it has not reached secular equilibrium. To compensate for the shorter emanation time, the following equation is applied:
\begin{equation}
    A(t_{\text{ema}}) = A^{\text{sec}} \times \left(1 - \exp(-\lambda_{\text{Rn}} t_{\text{ema}})\right),
    \label{eqn:blank_emanation}
\end{equation}
where  $A(t_{\text{ema}})$ is the activity resulting in an emanation time $t_{\text{ema}}$, and $A^{\text{sec}}$ is the activity at secular equilibrium. The emanation time in the measurements conducted is equivalent to the total time of gas system operation with the gas TPC, four days.

\autoref{tab:rgrresults} shows the extrapolated values for the radon activity at secular equilibrium without MS filtration, and the steady-state radon activity with MS filtration predicted by the radon dynamics model. The results demonstrate a clear reduction in the intrinsic radon activity in the TPC volume due to the application of the gas system containing low radioactive MS type 5\text{\AA}\ (NU MS V2).

\begin{table}[ht]
\centering
\begin{tabular}{|c|c|}
\hline
\textbf{Measurement Run} & \textbf{Extrapolated Steady State Activity (mBq)} \\
\hline
Without MS filtration  &  43.3$\pm$14.4 \\
With MS filtration      & 0.8$\pm$6.4 \\
\hline
\end{tabular}
\caption{\label{tab:rgrresults} Summary of radon activity results.}
\end{table}

The result for the measurement run without MS filtration is 43.3±14.4 mBq, representing the maximum radon activity during detector operation due to intrinsic material emanation from both the 100 L ThGEM-based TPC detector and the gas system prototype. For comparison, the larger DRIFT experiment, a 1.5 x 1.5 x 1.5 m\(^3\) gas-based directional dark matter detector, was measured to have a radon activity of 372±66 mBq \cite{Battat_2014}.

The application of MS filtration with low radioactive MS in the gas system prototype resulted in suppressing the radon activity to 0.8±6.4 mBq. The large error margin can be attributed to the background limits of the measurement apparatus. The MS filtration result before background subtraction was 14.8±2.8 mBq, which is within the error range of the background activity, 14.0±5.7 mBq. Thus, the radon activity has been reduced within the measurement limits of the apparatus. For a conservative calculation of the total activity reduction, the upper limit of the MS filtration result was used. Consequently, the gas system prototype utilising NU MS V2 has reduced the intrinsic radon activity in the ThGEM-based TPC detector setup to less than 7.2 mBq at a 95\% C.L., corresponding to a total reduction of at least 83\%.

\section{Gas gain conservation test}
\label{Chapter:GasSystemDemo:ggr}

To assess the performance of the gas system prototype in conserving gain in a ThGEM-based TPC detector, despite the presence of gain-harming common pollutants, the assessment was conducted over two runs. The first run was designed to monitor gain deterioration due to intrinsic contamination from the experimental setup. This was followed by a second run with the gas system loaded with \(3\text{\AA}\) and \(4\text{\AA}\) MS types, known to capture common pollutants \cite{breck1973zeolite}, and not to adsorb \ce{SF6} \cite{Ezeribe_2017}.

To highlight the extent of gain reduction, a significant initial gain is desired. Therefore, \ce{CF4} was selected as a proxy for \ce{SF6}, given its ability to attain superior gain. Despite the differences in target gases, \ce{CF4} is chemically similar to \ce{SF6} concerning the interaction with \(3\text{\AA}\) and \(4\text{\AA}\) MSs, as neither gas is adsorbed by these sieves.

To evaluate the performance of the gas system prototype in conserving gain due to the removal of common pollutants, it is necessary to have an experimental setup that monitors gain and is only affected by common pollutants. The signal gain of a detector is dependent on many parameters, namely the purity of the gas, amount of the gas and the avalanche electric field \cite{Guida_2020, sauli2004progress}. There are also temporary gain effects, such as charge-up, which alter the effective gain at the start of detector operation \cite{hauer2020study}. In order to observe the effect of only the common pollutants on gain, it is important to keep all these parameters constant and account for temporary effects.

The experimental setup previously detailed in \autoref{Chapter:GasSystemDemo:setup}, is configured so that the ThGEM-based TPC detector’s gain can be monitored as a function of common pollutants. During detector operation, the vessel pressure and high voltage supplied to the detector can be continuously logged to ensure that the amount of gas and the avalanche electric field are kept constant, respectively. Since the charge-up effect only occurs during the first few minutes of operation, if gain measurements are performed over the timescale of days, it can be ignored.

To monitor gain in the ThGEM-based TPC detector, a constant source of ionisation is required to provide the signal. An \ce{^55Fe} calibration source producing 5.89 keV x-rays was used to generate electron-ion pairs in \ce{CF4} in the TPC detector volume. To drift and amplify the electrons, high voltages in the ThGEM-based TPC were configured to settings known to provide a stable signal gain in 50 torr \ce{CF4} \cite{eldridge2021new, scarff2017developments}. High voltages of -855 V and 604 V were applied to the cathode and top of the ThGEM, respectively. The bottom of the ThGEM was grounded to the vacuum vessel. The amplified charges were detected and recorded using the electronics and DAQ described in \autoref{Chapter:GasSystemDemo:setup}. The recorded signals are in the form of a pulse height spectrum used to calculate gain.

To assess the performance of the gas system, the ThGEM-based TPC detector was operated for one week without the gas system to demonstrate the gain deterioration due to intrinsic detector contamination from common pollutants. This was followed by an identical detector run but with the gas system operating. For both runs, the gain was monitored by measuring the \ce{^55Fe} calibration source energy spectrum for a 5 minutes exposure every half hour. 

For the detector run with the gas system, the time between the gas system's swing and dilution process, \( t_{swg} \), was set to 24 hours, corresponding to a daily replacement of one detector volume. Since the rate of contamination from common pollutants is expected to be greater than the rate of radon emanation, the amount of MSs used was maximised. A total of 500g of \(3\text{\AA}\) and \(4\text{\AA}\) Sigma-Aldrich MSs in equal ratios were used for each filter.\\

The total gain of the TPC detector is a combination of the electronics gain, resulting from the amplification due to preamplifier and shaper electronics, and the gas gain, which arises from charge multiplication in the detector gas. Since this investigation focuses on the impact of common pollutants on charge multiplication, it is essential to calibrate the output signals from the ThGEM-based TPC detector to gas gain. The detector's response can be calibrated to correspond to gas gain by determining the expected charge from ionisation due to the calibration source and accounting for the contribution of the preamplifier. The expected charge from ionisation is a function of the energy of the \ce{^55Fe} calibration source, 5.8 KeV, and the average energy (\(W\)) required to create an electron-ion pair, 35 eV \cite{lopes1986ionisation}. The gain contribution from electronics comes from the preamplifier with a capacitance of 1.0 pF.  By simulating the ThGEM signal output with a range of known test pulses, it is possible to relate the pulse height spectra recorded by the Ortec 926 ADCAM analog-to-digital converter multichannel analyser to gas gain.

\begin{figure}[h]
    \centering
    \begin{minipage}{0.43\textwidth}
        \centering
        \includegraphics[width=0.85\linewidth]{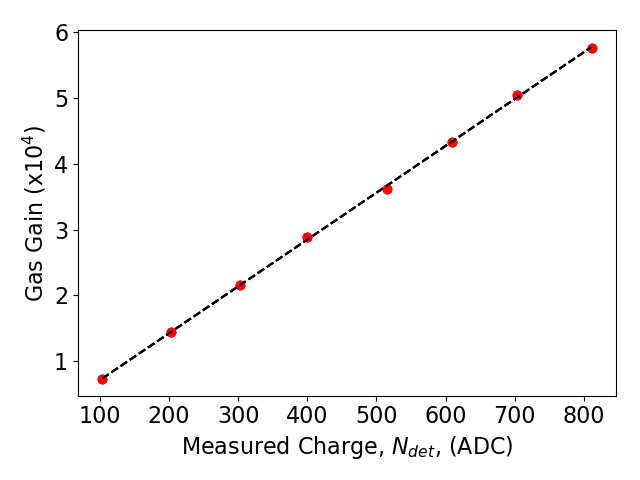}
        \caption{Plot of gas gain against ADC detector output.}
        \label{fig:gaincalplot}
    \end{minipage}\hfill
    \begin{minipage}{0.45\textwidth}
        \centering
        \includegraphics[width=0.8\linewidth]{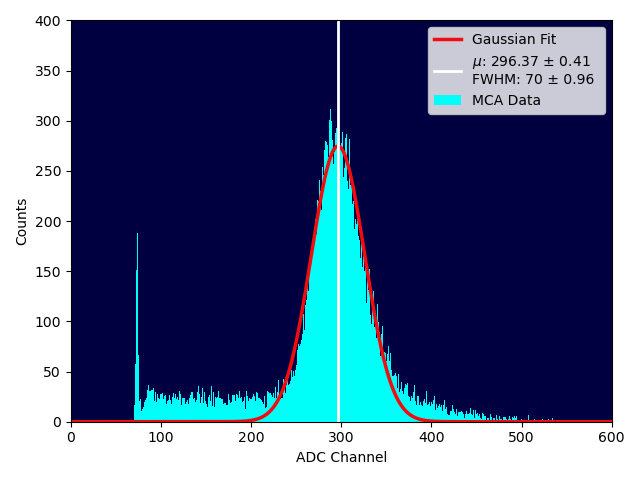}
        \caption{Gaussian fit to \ce{^55Fe} calibration source pulse height spectrum peak.}
        \label{fig:mcafit}
    \end{minipage}
\end{figure}

\autoref{fig:gaincalplot} illustrates the calculated gas gain from the test pulses ranging from 200 to 1600 mV, configured with Tennelec TC 814, plotted against the detector response (\(N_{\text{det}}\)).A least-squares regression fit provides the gas gain, \(GG\), calibration equation as follows:
\begin{align}
    GG &= 71.4 \times N_{\text{det}} - 105.
    \label{eqn:calibration}
\end{align}

\autoref{fig:mcafit} presents a measurement of the \ce{^{55}Fe} calibration source. This plot illustrates the number of counts for various pulse heights during a five-minute exposure to the \ce{^{55}Fe} source. Unlike the test pulses used in gain calibration, the pulse height spectrum obtained from the \ce{^{55}Fe} source is not as well-defined, primarily due to the inherent response characteristics of the ThGEM-based TPC detector \cite{knoll2010radiation}. To calculate the detector response (\(N_{\text{det}}\)), a Gaussian curve is fitted to the \ce{^{55}Fe} photo-peak signal in the pulse height spectrum, as indicated by the red line. The mean of this Gaussian fit (\(\mu\)), marked by the white line, is then determined. The gas gain is subsequently calculated using \autoref{eqn:calibration}, where the detector response (\(N_{\text{det}}\)) is equated to the Gaussian mean (\(\mu\)). \\

To determine gas gain variations over time, Gaussian analysis was applied to each measurement during the week-long detector run. It is worth noting that in the pulse height spectra, a decreasing gas gain corresponds to the \ce{^55Fe} photo peak signal shifting towards lower ADC channels. There is a period when the \ce{^55Fe} photo peak signal begins to leak into the ADC threshold. Beyond this point, the Gaussian analysis is no longer applicable, as the signal is lost to the background.

\begin{figure}[h]
\centering 
\includegraphics[width=6.5cm]{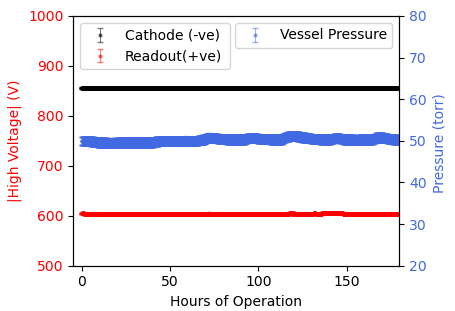}
\qquad
\includegraphics[width=6.5cm]{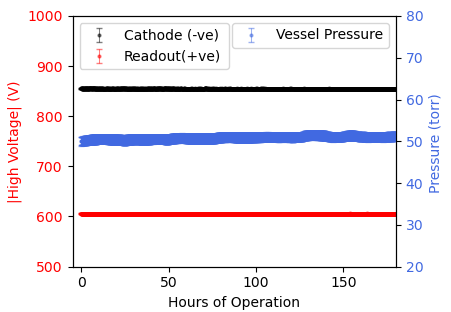}
\caption{\label{fig:otherparameters} Plot of detector pressure and applied high voltages in the ThGEM TPC during measurement runs. No gas replacement run (left) and gas system operation run (right).}
\end{figure}

Before analysing the data from the week long measurement runs, it is important to verify that other gain-affecting parameters were constant throughout. \autoref{fig:otherparameters} shows plots of detector pressure and high voltages applied to the ThGEM TPC, for both measurement runs. The average values and 2$\sigma$ deviation are shown in \autoref{tab:otherparameters}. Standard deviation in detector pressure and high voltages are within the instrumentation uncertainty of the logger. Consequently, any variations in detector gas gain during measurement runs can be attributed to the presence of common pollutants.

%

\begin{table}[h]
\centering
\begin{tabular}{|c|c|c|c|}
\hline
\textbf{Measurement Run} & \textbf{\begin{tabular}[c]{@{}c@{}}Average Detector \\ Pressure (Torr)\end{tabular}} & \textbf{\begin{tabular}[c]{@{}c@{}}Average \\ ThGEM HV (V)\end{tabular}} & \textbf{\begin{tabular}[c]{@{}c@{}}Average\\ Cathode HV (V)\end{tabular}} \\ \hline
Without Gas System  & 50.2 ± 0.8                                                                           & 604.0 ± 0.8                                                                & 855.0 ± 0.3                                                                 \\ 
With Gas System     & 50.9 ± 0.7                                                                           & 604.0 ± 0.1                                                                & 855.0 ± 0.6                                                                 \\ \hline
\end{tabular}
\caption{Average values of detector pressure and applied high voltages over the measurement runs. Errors shown are 2$\sigma$ deviation. }
\label{tab:otherparameters}
\end{table}

\begin{figure}[htb]
  \centering
  \begin{minipage}[b]{.47\textwidth}
    \centering
    \includegraphics[width=\linewidth]{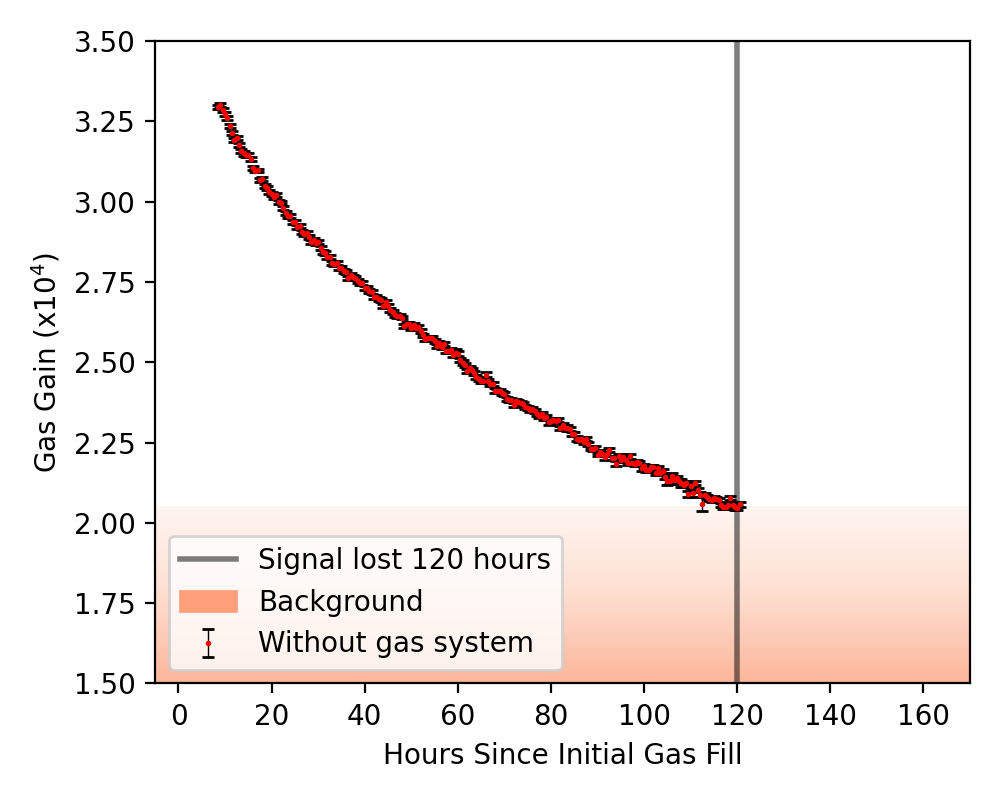}
    \caption{Plot of gas gain against time elapsed since initial gas fill for measurement run without gas replacement.}
    \label{fig:ggrwithoutgs}
  \end{minipage}\hfill
  \begin{minipage}[b]{.47\textwidth}
    \centering
    \includegraphics[width=\linewidth]{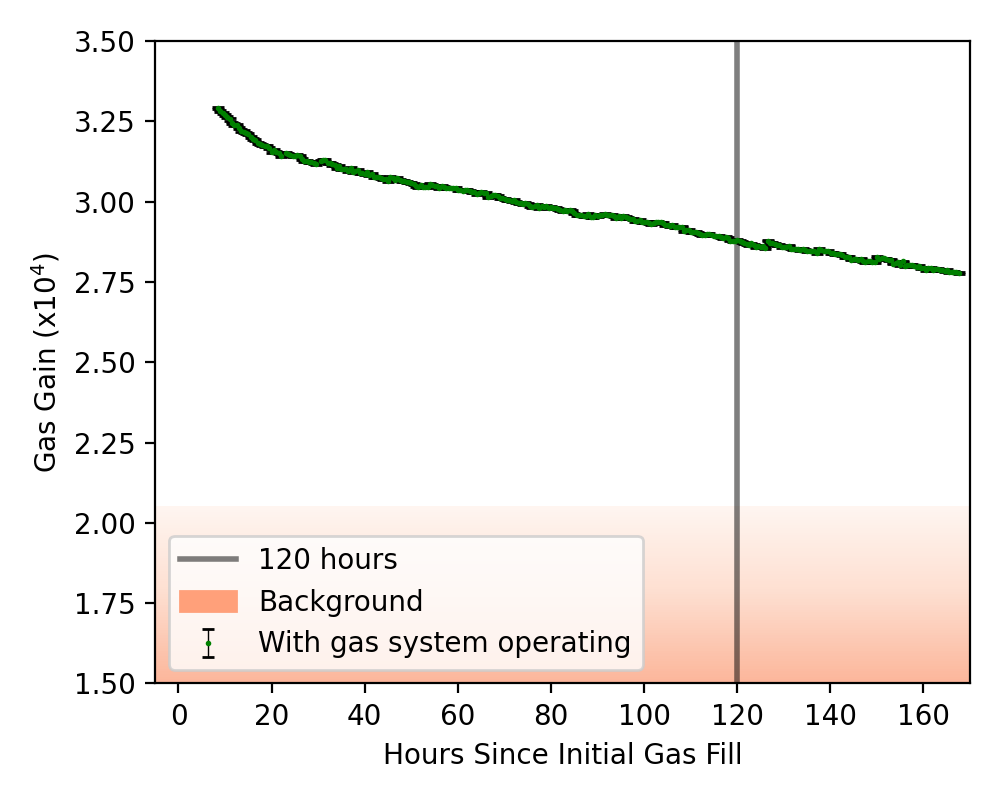}
    \caption{Plot of gas gain against time for measurement run with gas system operation. Scales are consistent with Fig. \ref{fig:ggrwithoutgs}.}
    \label{fig:ggrwithgs}
  \end{minipage}
\end{figure}

\autoref{fig:ggrwithoutgs} shows the ThGEM-based TPC’s gas gain over time without gas replacement. Although measurements were made for the full week, the gas gain could not be calculated from the pulse height spectrum after 120 hours as the \ce{^55Fe} photo peak started to leak into the ADC threshold background. Therefore, the signal is defined as lost at 120 hours, as indicated by the grey vertical line.
There is a clear deterioration of the gas gain over the week-long measurement, with a quarter of the initial measured gas gain lost after 54 hours since gas fill. The non-linear decrease in gas gain has been attributed to contaminants capturing both primary and avalanche electrons \cite{Guida_2020}.

\autoref{fig:ggrwithgs} presents gas gain measurements taken while the gas system prototype was operational. Throughout the measurement week, the gas gain consistently exceeded the background levels. In contrast, during the previous run without the gas system, the signal converged with the background noise after 120 hours (as marked by the grey vertical line). However, with the gas system in operation, the gain at the same interval was 87\% of the initial gain.

A notable feature of the gain plot with the gas system operating is the sudden decrease in gain deterioration. This can be explained by the gas system operation outlined in \autoref{Chapter:GasSystemDemo:setup}. Prior to the first gas dilution in the swing cycle, the TPC volume is comparable to a system operating without gas filtration, hence the initial gain reduction rate is very similar to the run without the gas system.

Furthermore, a distinctive feature of the gain plot is that upon gas dilution, the volume from the gas system, which has been undergoing filtration for a duration of \( t_{\text{swing}} \), is introduced. Ideally, a complete replacement of the TPC volume would reset the gas gain to initial levels. However, the process involves dilution—mixing the gas system volume with the TPC volume—indicating that the gas replacement is not entirely effective.

The first dilution resulted in a noticeable change in the rate of gain deterioration, indicating that the gas dilution process effectively removed gain-harming common pollutants from the TPC volume. The continued operation of the gas system appears to sustain the slowdown in gain deterioration. However, it is important to note that, despite this improvement, the gas gain is still gradually declining. This suggests that the rate of intrinsic contamination is greater than the filtration rate of the gas system. Consequently, it implies that the overall amount of common pollutants in the detector volume will continue to increase over time, until it reaches a critical contamination level, resulting in the loss of signal.

Simultaneously, as the concentration of common pollutants increases, more species become available for adsorption, potentially leading to an improved filtration rate. This suggests the possibility of reaching a steady-state gas gain where the contamination removed by the gas system equals the intrinsic contamination introduced between swing cycles. This concept is analogous to the steady-state radon activity predicted by the radon dynamic model in \autoref{Chapter:GasSystemDemo:rgr}. To investigate how the rate of gain deterioration evolves with gas system operation, the gain measurement run was extended for another week.

\begin{figure}[htb]
\centering
\begin{minipage}{.45\textwidth}
  \centering
  \includegraphics[width=\linewidth]{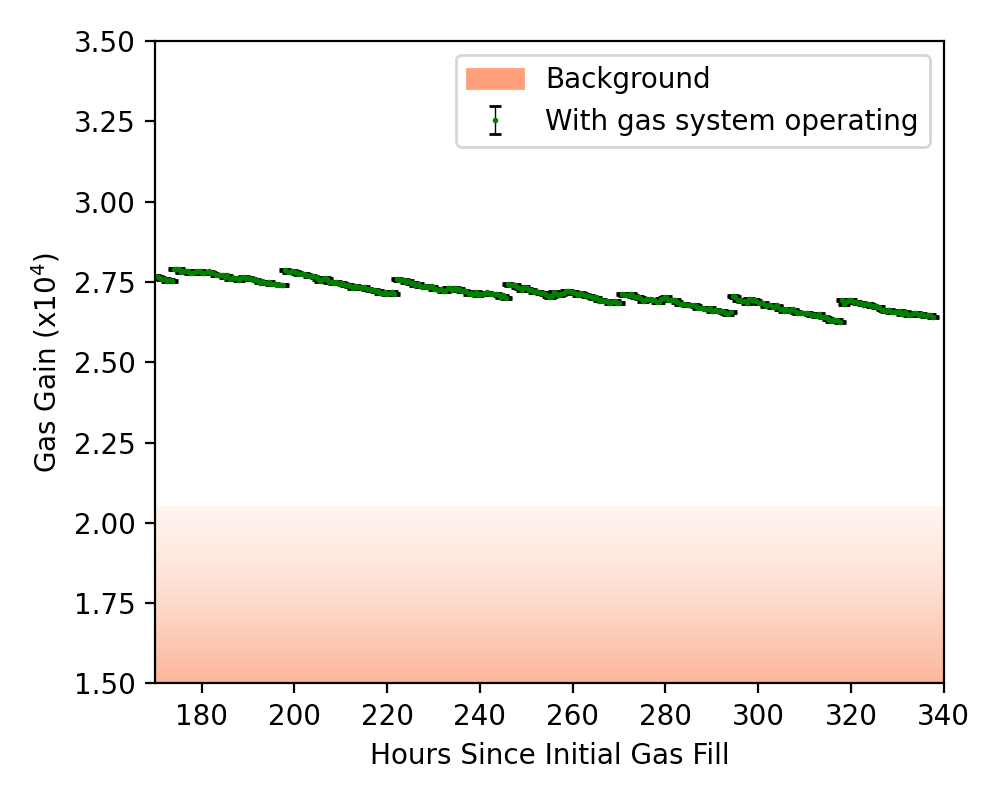}
  \caption{Plot of gas gain against time for extended measurement run with gas system operation. Scales are consistent with previous figures.}
  \label{fig:ggrwithgsext}
\end{minipage}\hfill
\begin{minipage}{.45\textwidth}
  \centering
  \includegraphics[width=\linewidth]{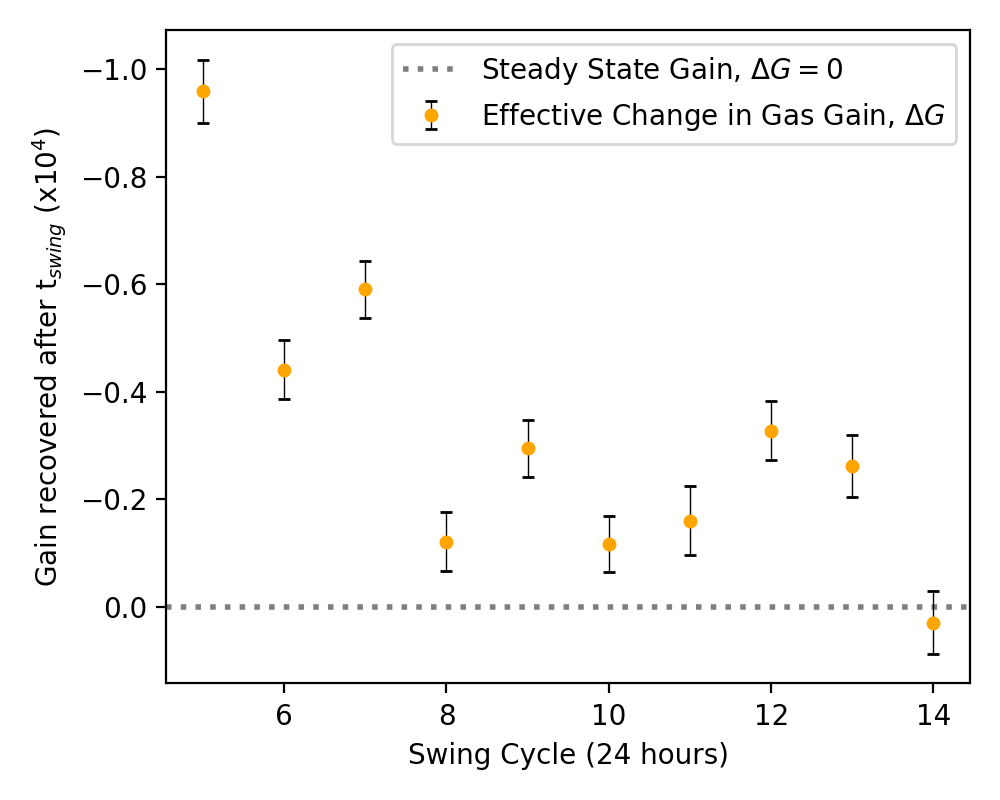}
  \caption{Plot of effective gain \(\Delta G\), against swing cycle. Note that y-axis is inverted.}
  \label{fig:deltag}
\end{minipage}
\end{figure}

Gain measurements with the gas system operating for another week are shown in \autoref{fig:ggrwithgsext}. The gas gain signal remained above the background until detector operation was stopped at 340 hours. The periodic discontinuity in gas gain corresponds to the gas dilution every 24 hours during the swing cycle. The detector pressure and high voltage monitors stayed within the observed deviation during the first week of measurements. 

The investigation into whether the gas system filtration and intrinsic contamination are moving towards a steady state involved analysing the effective gain change, $\Delta G$, following each swing cycle. Shown in \autoref{fig:deltag} is a plot of $\Delta G$ against the number of swing cycles. This change is defined by the equation
\begin{equation}
    \Delta G = G_{R} - G_{L},
\label{eqn:steadystategain}
\end{equation}
where $G_R$ represents the gain recovered, determined from the magnitude of the discontinuity following gas dilution, and $G_L$ is the gain lost, calculated from the difference between the initial and final gain within a swing cycle. A steady state gain would be implied when $\Delta G$ is zero.

There is a noticeable trend towards a decreasing magnitude of $\Delta G$ with each cycle. Notably, the final cycle exhibits a $\Delta G$ that is within the margin of error for a steady state condition of $\Delta G=0$. This trend indicates the potential for reaching a steady-state gas gain where the contamination removed by the gas system equals the intrinsic contamination introduced between swing cycles. To confirm the establishment of such an equilibrium, further data collection would be required. However, it is worth noting that during the final 48 hours of operation, the gas gain remained within the range observed during the last cycle, between 2.69 and 2.63 $\times10^4$.

To provide a fair comparison between the measurement runs, both the levels of gas gain maintained and the amount of gas used must be taken into consideration. A summary of the results from the measurement runs, which includes gas gain levels at notable points, is presented in \autoref{tab:ggrresults}. The percentage of gas gain is compared to the highest level measured in the detector setup, 3.3 $\times 10^4$. The 'Gas Used' corresponds to the total amount of gas utilised during operation, measured in units of TPC volume.

\begin{table}[h]
\centering
\begin{tabular}{|c|c|c|c|c|c|}
\hline
\multirow{2}{*}{\textbf{\begin{tabular}[c]{@{}c@{}}Measurement\\ Run\end{tabular}}} & \multirow{2}{*}{\textbf{\begin{tabular}[c]{@{}c@{}}Gas Used\\ (TPC vol.)\end{tabular}}} & \multicolumn{3}{c|}{\textbf{Gas Gain (\%)}} & \multirow{2}{*}{\textbf{Signal notes}} \\
& & \textbf{50h} & \textbf{120 h} & \textbf{340h} & \\
\hline
Without Gas System & 1 & 80\% & 62\% & - & lost after 120 hours \\
With Gas System & 2 & 92\% & 87\% & 80\% & remained until termination \\
\hline
\end{tabular}
\caption{Summary of the measurement runs results with gas gain levels at notable points.}
\label{tab:ggrresults}
\end{table}

In the run without gas replacement, the signal was lost to the background after 120 hours. In contrast, when the gas system was operating, the signal remained above the background until detector operation was stopped at 340 hours. One might argue that the gas system, requiring two TPC volumes during operation, used double the amount of gas. If the same amount of gas were used in the measurement without the gas system, the run would extend to 240 hours. However, it is essential to consider the sustained levels of gas gain in this comparison.

In the measurement run with the gas system, the gas gain remained above $2.63 \times 10^4$, which is $80\%$ of the highest gain achieved in the setup, for $340$ hours. In contrast, without the gas system, the gas gain only stayed above this level for $50$ hours. Assuming the same gain deterioration rates, to maintain a gas gain of at least $80\%$ without the gas system for $340$ hours, the detector volume would need to be replaced seven times, requiring $3.5$ times more gas compared to the detector run with the gas system operating.

The gas gain remained above the background level until the measurement run was terminated, indicating the potential for extended operation beyond 340 hours. Furthermore, during the last swing cycle, the amount of gain recovered and gain lost were within errors, which raises the possibility of achieving a steady-state gain. The last two swing cycles resulted in values equivalent to 81±1\% of the highest gain measured in the setup.

\section{Conclusions}

In this paper, we described the tests conducted to evaluate the concept of an MS-based vacuum swing adsorption gas system design. A prototype gas system was applied to a ThGEM-based TPC detector to assess its ability to reduce intrinsic radon contamination from the detector setup and maintain detector gain by removing gain-harming common pollutants. It was shown that when coupled with the low radioactive MS type \(5\text{\AA}\) (NU MS V2), the gas system prototype effectively reduced the intrinsic radon activity in the ThGEM-based TPC detector setup within the margin of error of the radon measurement apparatus background (\(14.0 \pm 5.7\) mBq). Using the upper limits of radon measurement, we determined that radon activity had been reduced to less than \(7.2\) mBq at a 95\% C.L., corresponding to a reduction of at least \(83\%\) of the total intrinsic radon activity of the setup.

Additionally, it was demonstrated that utilising MS types \(3\text{\AA}\) and \(4\text{\AA}\) with the gas system significantly mitigated the effects of gain deterioration due to common pollutants. In a detector run with the gas system operating, the signal remained until detector operation was terminated after 340 hours. In contrast, without the gas system, the TPC detector could only maintain this level of signal amplification for 50 hours. Furthermore, an extended detector run with the gas system suggests that a steady-state gain, where the introduction of common pollutants equals the filtration, is potentially attainable. However, the implied steady-state gain is \(80\%\) compared to the gain with fresh gas.

The results presented in this paper suggest that a vacuum swing adsorption gas recycling system, when coupled with suitable MSs, has the potential to significantly reduce intrinsic radon activity and extend detector operation in an \ce{SF6} gas-based directional dark matter detector. This capability is crucial not only for the current experimental setup in development but also for the successful operation of future large-scale ultra-sensitive gas-based searches, such as CYGNUS-1000 \cite{vahsen2020cygnus}. However, further research is required to optimise the operation and components of the gas system, which can enhance both the reduction of radon levels and the conservation of gas gain. This includes fine-tuning filtration by determining the ideal gas system parameters, such as flow rate, pressure, temperature, and \(t_{\text{swg}}\). Additionally, reducing intrinsic radon and common pollutant contamination can be achieved through extensive screening of the gas system components used.

\acknowledgments

The authors would like to acknowledge support for this work through the EPSRC Industrial CASE with DURRIDGE UK Award Grant (EP/R513313/1)


\bibliographystyle{JHEP}  
\bibliography{references}  

\end{document}